\documentclass[traditabstract]{aa}
\usepackage{amssymb,amsmath,amsfonts}
\usepackage{graphicx,graphics}
\usepackage[usenames]{color}
\usepackage{ulem}

\newcommand{\beq}{\begin{equation}}
\newcommand{\eeq}{\end{equation}}
\newcommand{\bea}{\begin{eqnarray}}
\newcommand{\eea}{\end{eqnarray}}

\newcommand{\bhl}{}
\newcommand{\ehl }{}

\begin{document}
\title{The growth of helium burning cores}
\titlerunning{to be set}

\author{H.C.\ Spruit$^{\small 1,2}$}
\authorrunning{H.C.\ Spruit}

\institute{
  Max-Planck-Institut f\"{u}r Astrophysik,
  Karl-Schwarzschild-Str.\ 1,
  D-85748 Garching, Germany \label{inst1}
  \and
 Monash Center for Astrophysics, Monash University, VIC 3800, Australia\label{inst2}
}

\date{\today}

\abstract{
Helium burning in the convective cores of horizontal branch and `red clump' stars appears to involve a process of `ingestion' of unburnt helium into the core, the physics of which has not been identified yet. I show here that a limiting factor controlling the growth is the buoyancy of helium entering the denser C+O core. It  yields a growth rate which scales directly with the convective luminosity of the core, and agrees with constraints on core size from current asteroseismology. 

 
}

\maketitle

\section{Introduction}

\medskip

After their life as hydrogen shell burning stars on the red giant branch (RGB), stars in a wide mass range settle as core helium burning stars on the horizontal branch (HB).  The burning core is convective due to the large temperature sensitivity of the helium burning reaction. After the helium in the core is exhausted, He-burning continues in a shell around the resulting carbon+oxygen core. In this stage the stars populate the asymptotic giant branch (AGB).  Observations of the relative frequency of HB and AGB stars can be used to constrain the mass of He that has been burnt at the point when He shell burning sets in. Recent asteroseismic evidence shows this mass to be significantly larger than can be explained with basic stellar evolution (Constantino et al. 2015): it is necessary to assume some process of `convective overshooting', or `ingestion' of He from the convectively stable region outside into the burning core. Various parametrization are used to quantify such ingestion, and tuned to match the resulting evolution paths (cf.\ Salaris \& and Cassisi 2005,  Gabriel et al.\ 2014, Constantino et al. 2015). Inspection of the hydrodynamics involved, however, reveals a piece of physics that appears to have been missed in previous discussions. It puts important an constraint on the rate of ingestion  to be satisfied by any parametrization. 

He being highly buoyant relative to the He-C-O mixture in the core, only a very small concentration of additional He can be carried down in the convective flows. The  convective heat flux in the core is carried by a temperature fluctuation $\Delta T$ of only  $\approx 1$ K, or $\Delta T/T\approx 10^{-8}$.  
An upper limit to the rate at which He can enter the core is given by assuming that the  convective downflows carry a He excess that just keeps the buoyancy of the flow negative (downward).  This limit corresponds to density contrast of the same order as the convective temperature contrast. Since the downward mass flux is the same as the upward mass flux carrying the (convective) luminosity of the burning core, this limit translates into a direct relation between He ingestion rate and luminosity. In the following it is shown that the actual rate is probably not too far below this upper limit. This makes it a relevant and convenient quantity for parametrizing ingestion. 

\section{Ingestion}
The convective heat flux (erg cm$^{-2}$s$^{-1}$) is
\beq F_{\rm c}=v\rho c_{\rm p} \Delta T,\eeq
where $v$  and $\Delta T$ are suitable averages (from a mixing length formalism, say) of  the convective flow speed and the temperature contrast associated with it, and $c_{\rm p}$ the specific heat per unit mass at constant pressure. To illustrate the idea assume first a simple, fully ionized ideal gas equation of state, such that $P={\cal R}\rho T/\mu$, and $c_{\rm p}={5/2\,\cal R}/\mu$, where $\cal R$ is the gas constant and $\mu$ the mean weight per particle in atomic units. In pressure equilibrium, this yields, for small changes $\delta$:
\beq \delta\ln\rho+\delta\ln T-\delta \ln\mu=0.\eeq
Setting  $\delta\rho=0$ yields an upper limit to the ingestion rate. Equating $\delta\ln T$ with the convective amplitude $\Delta\ln T$, and with $\delta\ln\mu$ being of the order of the change in mass faction $\delta x$ of the He entering, the downward mass flux of excess He has a maximum of order
\beq 
\dot m_{\rm i}= v\rho\,\delta x\approx v\rho\,\delta\ln\mu<v\rho\,\Delta\ln T={F_{\rm c}\over c_{\rm p}T},\label{dotm0}\eeq
where $\dot m_{\rm i}$ is the downward mass flux (g cm$^{-2}$ s$^{-1}$) of He ingested. This shows the main result: buoyancy of the lighter component limits the ingestion rate to a value that is directly related to the convective heat flux.  With a factor $4\pi r^2$, (\ref{dotm0}) gives the total He mass ingestion rate $\dot M_{\rm i}$:
\beq
\dot M_{\rm i}\sim\alpha_{\rm i} {L_{\rm cc}\over c_{\rm p}T},\label{mdoting}\eeq
where $L_{\rm cc}$ is the fraction of the core luminosity $L_{\rm c}$ that is carried by convection, and $\alpha_{\rm i}$ is a dimensionless {\it ingestion efficiency} parameter, the value of which is still to be determined. 

At the maximum set by assuming neutral buoyancy, the density deficit due to the additional He  just compensates the density excess of the cool downward flow, so this addition actually would not flow down. The opposite limit of vanishing He contrast does not carry any additional He either. The optimum is probably somewhere in the middle. In the absence of physics for a quantitative estimate how much the He actually makes it into the downward plumes, a reasonable upper limit to $\alpha_{\rm i}$ is `halfway between':
\beq \alpha_{\rm i} < 0.5. \label{uplim}\eeq

Eq.~(\ref{mdoting}) is a rather  coarse estimate.  For a more general equation of state $P(\rho,T,\mu)$, the neutrally buoyant contrast in $\mu$  at constant pressure would be given by
\beq (\delta\ln\mu)_{\rm max}= \mu_T\,\Delta\ln T;\quad {\rm with }\quad\mu_T\equiv \left({\partial\ln \mu\over\partial\ln T}\right)_{P,\rho}\eeq
If $x$ is the mass fraction of helium, and 
\beq \mu_x\equiv{\partial\ln \mu\over\partial x}\vert_{P,T}\,,\eeq
we have
\beq (\delta x)_{\rm max}=(\delta\ln\mu)_{\rm max}/\mu_x={\mu_T\over \mu_x}\,\Delta\ln T. \eeq

For an ideal gas equation of state, $\mu_T=1$. For elements of masses $m_1,\,m_2$ (in atomic units) and degrees of ionization $q_{1,2}$,  the mass fraction $x_1=\rho_1/\rho$ is related to $\mu$ by
\beq 
\mu^{-1}=x_1(r_1-r_2)+r_2; \quad{\rm with~~~} r_{1,2}={1+q_{1,2}\over m_{1,2}},\eeq
so that 
\beq \mu_x=\mu (r_2-r_1)\label{dmux},\eeq and
eq.\  (\ref{dotm0}) is replaced by
\beq\dot M_1= \alpha_{\rm i}{\mu_T\over\mu_x}{L_{\rm cc}\over c_{\rm p}T},\label{est1}\eeq
(counted positive for downflow). For a fully ionized He-C mix $m_1=4$, $q_1=2$, $m_2=12$, $q_2=6$,  $r_2-r_1 =1/6$, for example.    With these numbers eqs.\ (\ref{est1}) and (\ref{dmux}) then yield an improved estimate, using  $\mu c_{\rm p}=5/2\,{\cal R}$ for a fully ionized ideal gas :
\beq\dot M_{\rm i}\approx \alpha_{\rm i}{12\over 5}{L_{\rm cc}\over{\cal R}T}\label{estHe}.\eeq

\subsection{Ingestion vs. mixing}
Expression (\ref{est1}) gives the rate at which the convective core `eats its way into' the overlying stable layer. In this view, ingestion and growth of the convective core are the same thing\footnote{At least as long as secular changes in the structure of the \bhl star\ehl can be neglected.  Associated with the slow contraction of the core the radiative gradient increases, causing the mass coordinate of the convective boundary to move out even in the absence of any mixing or ingestion.}.  Descriptions of ingestion in terms of convective `overshooting' do not quite capture the essence: the buoyancy constraint. 

The composition jump between the CO-rich core and the helium above also plays an important role. At a convective boundary between zones of identical composition, mixing  or `overshooting' is facilitated by the high efficiency by which thermal diffusion reduces the stabilizing buoyancy on the radiative side of the boundary. This is not the case if the stability of the boundary is due to a difference in composition.  In this case, the extent of mixing of the burning product (C+O) into the overlying He, for example by shear flow instabilities, is limited by the low convective flow speeds driving the instability. Richardson's condition for shear instability  for example (governed by the compositional buoyancy), would be met only in a very thin boundary layer. 
The parallel process of  He mixing into the CO core is also slow, limited by the buoyancy of He in the C-O-He mixture of the core, but its consequences are more interesting. The He that gets mixed in (making the core grow in mass) is quickly carried down to the burning region of the core. 

The core's convective luminosity $L_{\rm cc}$ is less than the star's entire luminosity, since  the H-burning shell  and some gravitational luminosity also contribute. More significant, however, is a contribution from radiative transport. Above the boundary the entire core luminosity is carried  by radiation. Associated with  the difference in composition there is a jump in opacity at the convective boundary, but not by a large factor.  This implies that radiative transport must also be significant below the boundary.  $L_{\rm cc}$  is correspondingly smaller than the luminosity of the core, reducing our estimate of the ingestion rate (\ref{estHe}). 

\subsection{Ingestion vs. burning}

The rate of ingestion of He can be compared with the rate at which it is burned. To this end, consider the energy balance of the core:
\beq
L_{\rm b}+L_{\rm g}=L_{\rm cc}+L_{\rm cr},
\eeq
where $L_{\rm b}$ is the luminosity represented by triple-$\alpha$ burning, and $L_{\rm g}$ the gravitational luminosity due to contraction of the core. Their sum is balanced by $L_{\rm cc}$ and $L_{\rm cr}$,  the convective and radiative luminosities at the core boundary. The core temperature is kept close to constant by the high temperature sensitivity of the burning process, but the increase in mean weight per particle from He to C+O causes the core to contract slightly, providing a gravitational energy source. Since the burning does not change the mean weight per particle much, however, this is probably a  modest contribution. 

With (\ref{estHe}), the ratio of  the ingestion and burning rates is
\beq
{\dot M_{\rm i}\over\dot M_{\rm b}}=\alpha_{\rm i}{12\over 5}\,{\epsilon_{\rm b}\over {\cal R} T}{L_{\rm cc}\over  L_{\rm cc}+L_{\rm cr}-L_{\rm g}},\label{ratio}
\eeq
where $\epsilon_{\rm b}$ is the energy released per unit mass of He burned. Making use of the definition of the radiative gradient $\nabla_{\rm r}$ we have 
\beq {\nabla_{\rm cr}\over\nabla_{\rm c}}={L_{\rm c}\over L_{\rm cr}}=1+{L_{\rm cc}\over L_{\rm cr}},\eeq
where the subscript $_{\rm c}$ refers to conditions  in the outer part of the convective core and $\nabla_{\rm c}$ is the logarithmic temperature gradient, nearly equal to its adiabatic value $\nabla_{\rm ca}$. Omitting the small contribution from $L_{\rm g}$, eq.\ (\ref{ratio}) yields
\beq
{\dot M_{\rm i}\over\dot M_{\rm b}}=\alpha_{\rm i}{12\over 5}\,{\epsilon_{\rm b}\over {\cal R} T}(1-{\nabla_{\rm ca}\over\nabla_{\rm cr}}).\label{ratio1}
\eeq

\subsection{Estimate of ingestion efficiency}

If He burns faster than it is ingested ($\dot M_{\rm b}>\dot M_{\rm i}$),  its concentration in the convective core will decrease. If this is the case throughout the helium burning phase, helium in the convective core is exhausted before all of it has been ingested. The star then develops a helium burning shell and becomes an asymptotic giant. 

In the opposite case  $\dot M_{\rm i}>\dot M_{\rm b}$, the burning core would grow until all helium available has been incorporated into it. There would be no helium shell phase, contrary to what is inferred from observation. 
The fact that  both the AGB and the horizontal branch are significantly populated therefore tells us that $\dot M_{\rm i}/\dot M_{\rm b}$ appears to be somewhat smaller than unity, maybe of order 0.5 but not much less. This allows us to make an estimate of the ingestion efficiency $\alpha_{\rm i}$. 
A $1M_\odot$ model (kindly provided by Simon Campbell) has  $\nabla_{\rm ca}/\nabla_{\rm cr}\approx 0.95$ in the outer regions of the growing core (see also figs. 6.4-6.6 in Salaris and Cassisi 2005). Most of the luminosity is actually carried by radiation there, and the ingestion rate \bhl deduced\ehl is only about 5\% of what it would be if the energy flux were fully carried by convection. With the observational clue implying $\dot M_{\rm i}/\dot M_{\rm b}\approx 0.5$, with $\epsilon_{\rm b}=6\,10^{17}$ erg/g, a temperature at the convective boundary of about $ 9\,10^7$ K, and $\mu c_{\rm p}\approx 5/2\,{\cal R}\approx 2\, 10^8$ erg g$^{-1}$ K$^{-1}$, eq.\ (\ref{ratio1}) then yields 
\beq \alpha_{\rm i}\approx 0.1\label{est2}\eeq
as our  provisional estimate of the ingestion efficiency in core helium burning, i.e. 20\% of the upper limit (\ref{uplim}).

\subsection{Structure of the core boundary}
The jump in mean weight per particle presents an obstacle to any mixing process that could potentially broaden the gradient composition at the core boundary. At the convective flow speeds in the core,  the shear across the $\mu-$jump would have  to be concentrated in a layer of  the order of a meter in order to satisfy the Richardson criterion for  shear instability, for example. The stable $\mu-$jump supports surface waves (internal gravity waves), which can propagate some distance into the stably stratified region outside the core, but need not lead to much mixing. 

\bhl  The hydrodynamics at a convective boundary is becoming amenable to increasingly realistic numerical simulations.  Recent results by Arnett et al.\ 2015 address the core overshooting problem, with or without the effects of buoyancy at a jump in composition, from a theoretical turbulence perspective. A comparison of the resulting overshooting effects with observations is given in Schindler et al.\ (2015). Woodward et al.\ (2015) address  hydrogen ingestion into a He burning core undergoing a He-shell flash, using  direct numerical simulations tailored to this case. 

Details of the structure of a convective boundary as seen in simulations may well depend on the structure of the evolutionary state of interest. For the case of convection at a jump in composition, however, I expect the estimate (\ref{ratio1}) to be a fairly stable one,\ehl since ingestion into a convective burning core is limited to fluid elements with a positive density contrast. \bhl This makes it\ehl  independent of the mixing processes that determine the width of the composition gradient at the boundary. 

\section{Conclusion and discussion}
The strong buoyancy of He entering in a C+O+He mix limits the rate at which He can be `ingested' into the core. The positive density contrast driving the convective downflows, of the order $\Delta\rho/\rho=10^{-8}$, is offset by buoyancy of ingested helium at a contrast of this same order. This sets an upper limit on any mixing process taking place at the convective boundary. The actual ingestion rate is predicted to be significant fraction of this limit.

The transition region at the convective boundary may well be smoothened by shear instabilities or internal wave flows. Only the portion of this region that satisfies the buoyancy constraint, however, will actually be carried down with the convective flows. The width of the transition therefore has little influence on the ingestion rate. \bhl Whether or it is significant, most of the width\ehl  is too buoyant to enter the convective core. 

On the other hand, fluid with a helium excess satisfying the constraint is likely to be carried down into the core. The limit therefore also provides a good basis for estimating the actual ingestion rate. This provides a simple and direct relation between the ingestion rate and the convective luminosity of the core (eq.\ \ref{est1}). 

The convective luminosity is lowest near the core boundary, since this is where the transition to radiative transport takes place. The value of the ingestion rate is therefore set in the region just below the core boundary.

The estimate (\ref{ratio1}) depends somewhat on the rather simplistic approximation of an  ideal gas. This can easily be improved with the more realistic equations of state used in stellar evolution of core He-burning stars.  

Seismology made possible with data from the Kepler and Corot satellites is currently yielding constraints on size of the core in He-burning stars (Constantino et al. 2015, Bossini et al. 2015). These should make it possible to substantially improve the calibration of the ingestion efficiency factor estimated here (eq.\ \ref{est2}). 

It would be interesting to see if evolution calculated with ingestion as formulated here does or does not lead to the somewhat controversial `core breathing pulses', which occur  towards the end of core He-burning in some calculations (e.g.\ Castellani et al. 1985, Sweigart et al. 2000, Salaris \& Cassisi 2005).

\begin{acknowledgements} 
The work reported was  initiated during a visit at the Monash Center for Astrophysics.  The author thanks  MoCA for its kind hospitality, Alex Heger, John Lattanzio, Simon Campbell and Thomas Constantino  for discussions that  triggered the work, \bhl and David Arnett for input at the refereeing stage.\ehl The text also benefitted from error detection and  detailed comments by Achim Weiss.
\end{acknowledgements}


\begin{thebibliography}{99}

\bibitem{} Arnett, W.~D., Meakin, C., Viallet, M., et al.\ 2015, \apj, 809, 30 

\bibitem{} Schindler, J.-T., Green, E.~M., \& Arnett, W.~D.\ 2015, \apj, 806, 178 

\bibitem{} Bossini, D., Miglio, A., Salaris, M., et al.\ 2015, arXiv:1507.07797 

\bibitem{} Castellani, V., Chieffi, A., Tornamb\`e, A., \& Pulone, L.\ 1985, \apj, 296, 204 

\bibitem{} Constantino, T., Campbell, S.W., Christensen-Dalsgaard, J., et al. 2015, arXiv:1506.01209v1

\bibitem{} Gabriel, M., Noels, A., Montalb{\'a}n, J., \& Miglio, A.\ 2014, \aap, 569, A63 

\bibitem{} Salaris, M., \& Cassisi, S.\ 2005, Evolution of Stars and Stellar Populations, sections 6.2-6.4. ISBN 0-470-09220-3, Wiley-VCH.

\bibitem{} Sweigart, A.~V., Lattanzio, J.~C., Gray, J.~P.,  \& Tout, C.~A.\ 2000, Liege International Astrophysical Colloquia, 35, 529 

\bibitem{} Woodward, P.~R., Herwig, F., \& Lin, P.-H.\ 2015, \apj, 798, 49 






\end{thebibliography}
\end{document}